\documentclass{aa}
									      
\usepackage{txfonts,graphicx}															
\begin{document}

\title{Automated Nonlinear Stellar Pulsation Calculations: \ Applications to RR Lyrae stars}
\subtitle{The Slope of the Fundamental Blue Edge and the First RRd Model Survey}

\titlerunning{Automated Nonlinear Stellar Pulsation Calculations}
 
  \author{R. Szab\'o
          \inst{1}
	  \and
	  Z. Koll\'ath
	  \inst{1}
	  \and
	  J.~R. Buchler
	  \inst{2}}

  \offprints{R. Szab\'o \email{rszabo{\rm\char'100}konkoly.hu}}

   \institute{Konkoly Observatory, H-1525, Budapest, P.O.Box 67. Hungary
	    \and
	      Physics Department, University of Florida, Gainesville, FL 32611, USA
             }

   \date{Received 14 November 2003 / accepted 11 June 2004}

   \abstract{We describe a methodology that allows us to follow the pulsational
behavior of an RR~Lyrae model consistently and automatically along its
evolutionary track throughout the whole instability strip.  It is based on the
powerful amplitude equation formalism, and resorts to  a judicious combination of
numerical hydrodynamical simulations, the analytical signal time-series
analysis, and amplitude equations.

\noindent A large-scale survey of the nonlinear pulsations in RR~Lyr instability
strip is then presented, and the mode selection mechanism is delineated
throughout the relevant regions of parameter space.  We obtain and examine two
regions with hysteresis, where the pulsational state depends on the direction of
the evolutionary tracks, namely a region with either fundamental (RRab) or
first overtone (RRc) pulsations and a region with either fundamental (RRab) or
double-mode (RRd) pulsations.  

\noindent The regions where stable double-mode (DM, or RRd) pulsations occur
are very narrow and hard to find in astrophysical parameter ($L$, $M$, ${T_{\rm
ef\!f} }$, $X$, $Z$) space with hydrodynamic simulations, but our systematic
and efficient methodology allows us to investigate them with unprecedented
detail.

\noindent It is shown that by simultaneously considering the effects of mode
selection and of horizontal branch evolution we can naturally solve one of the
extant puzzles involving the topologies of the theoretical and observed
instability strips, namely the slope of the fundamental blue edge.  

\noindent The importance of the interplay between mode selection and stellar
evolutionary effects is also demonstrated for the properties of double-mode
RR~Lyr. Finally, the Petersen diagram of double-mode RR~Lyr models is discussed
for the first time.

}      
\maketitle

%======================================================================

\section{Introduction}

The evolutionary tracks of the classical variable stars carry them once or more
through the instability strip (IS) where they can undergo pulsations of various
types, such as single-mode pulsations in the fundamental (F), in the first
overtone (O1), in the second overtone (O2), or double-mode (DM) pulsations.
The latter are beat pulsations that occur either in the F and O1 modes or in
the O1 and O2 modes.  A beautiful
observational summary for the Cepheids is provided by the microlensing
projects, {{\it e.g.}\ } Beaulieu~{\it et~al.} (\cite{beau97}) or Udalski~{\it
et~al.}  (\cite{Udalski}).  Thus low luminosity Cepheids are seen to undergo DM
pulsations, whereas this phenomenon does not occur for the high luminosity
ones.  In the intermediate luminosity range single-mode F and O1 Cepheids seem
to overlap, and in the high luminosity regime there are only F Cepheids.  For
RR~Lyrae the picture is similar, except that there is another type of modulated
pulsation (the Blazhko effect) that has not received a satisfactory explanation
to date.

Theory of course aims to understand the observed modal behavior and to provide
a global picture thereof. In this paper we merge several techniques that have
been used separately and combine them into an efficient methodology that allows one to
follow the pulsational behavior of a star along its full evolutionary track,
from the moment it enters the IS till it leaves it.

These tools are then applied to open questions and relatively unexplored topics
concerning the radial pulsations of RR~Lyr stars. Since the first claim of a
successful modelling of double-mode pulsation with a turbulent convective code
(Feuchtinger \cite{feuchtinger1}, Koll\'ath~{\it et~al.}  \cite{kollath3}), no
systematic exploration of the RRd models has been published except for some
limited studies (Szab\'o~{\it et~al.}  \cite{rszabo}).  Here we attempt to
remedy this situation.  In the process of studying double-mode RR~Lyr we have
also uncovered a hidden relation between evolution and pulsation.

Another question concerns the discrepant topologies of the theoretical and the
empirical RR~Lyr ISs (Jurcsik \cite{jurcsik97}).  A series of
papers (Kov\'acs \& Jurcsik \cite{kj1}, \cite{kj2}; Jurcsik \cite{jurcsik})
establishes \emph{empirical relations} between RRab light-curve parameters and
absolute magnitudes, colors and metallicities.  These quantities are further
transformed to the ${T_{\rm ef\!f} } - L$ plane {\it via\ } Kurucz's static
model atmospheres.  (We mention in passing that such empirical relations do not
yet exist for the overtone RRc stars.)  For the distribution of RRab stars the empirical
${\rm Log}\thinspace {T_{\rm ef\!f} } - {\rm Log}\thinspace L$ diagram shows a 
well-defined and straight edge.  However, attempts at reproducing the empirical 
RR~Lyr F and O1 blue edges have failed: neither radiative nor convective models 
could explain the shallow slope of the blue edges of the F and O1 instability regions
(Koll\'ath~{\it et~al.} \cite{kbf}, KBF).  Despite the investigation of a large
number of physical effects that might change the slopes, such as rotation,
exotic chemical abundances, radiative transport, KBF could not offer an
explanation for the shallow slope of the empirical F blue edge.  Throughout
this paper we shall refer to this problem as the \emph{fundamental blue edge
problem}.  Using our new methodology, we will demonstrate that the solution to the
problem lies in the simultaneous taking into account \emph{both mode selection
and stellar evolution}.

Because of an ambiguous use of the notion of DM or beat pulsations in the
literature, at the risk of being pedantic, we need to address this issue
briefly.  We talk about {\sl DM pulsations} of a given {\sl stellar model},
with fixed $L$, $M$ and {$T_{\rm ef\!f}$} when (a) the Fourier 
spectrum exhibits two dominant frequency peaks and those corresponding to all the 
linear combinations of these two, and (b) the amplitudes of all the peaks are 
constant. When the pulsations of a stellar model display this type of spectrum, 
but the amplitudes vary slowly compared to the period (but still fast compared
to evolution), then such a model is in the process of 
{\sl switching modes} -- it may or may not ultimately end up as a true DM pulsator. 
In fact, it may just be switching from one single mode state to another, say F to 
O1, if it is in a so-called either F or O1 region.  But these different scenarios 
correspond to stellar models with different physical parameters.  For a 
{\sl real star}, $L$ and {$T_{\rm ef\!f}$}\ are not constant, but change as the 
star moves along its evolutionary track.  How this evolution affects mode switching 
has recently been discussed by Buchler \& Koll\'ath (\cite{bk02}, hereafter BK2002). 
From an observational point of view it may not be possible to discriminate between
stars that are switching from RRab to RRc (or RRc to RRab) and actual RRd
stars, but at least, in principle, they should be distinguished.

The layout of the paper is the following: the methodology is presented in
Sec.~{\ref{sec2}}., and it is applied in a large-scale nonlinear survey of 
radial RR~Lyr pulsation in Sec.~{\ref{sec3}}. In Sec.~{\ref{sec4}}. mode selection 
characteristics are discussed including the systematic mapping of the DM
pulsation region of RR~Lyr stars.  
The F blue edge problem is revisited in Sec.~{\ref{sec5}}., 
testing the combined effects of the \emph{mode selection} and horizontal-branch (HB) 
stellar evolution on the structure of the
RR~Lyr IS.  Three different sets of HB evolutionary tracks are
considered to account for realistic evolutionary changes of stellar parameters.
In Sec.~{\ref{sec6}}. an hitherto unknown connection is uncovered 
between double-mode pulsation and evolution, as well as the Petersen diagram of
nonlinear RRd models is discussed.  Finally, we conclude in Sec.~{\ref{sec7}}.

%=======================================================================

\section{The Methodology}
\label{sec2}

\subsection{Timescales and Amplitude Equations (AEs)}

Stellar pulsations are characterized by three timescales, generally well
separated in size. The shortest is the dynamic time, $\tau_{dyn}$, which
corresponds to the period or cycling time.  Next comes the transient time
scale, $\tau_{trans}$, on which the amplitudes change during mode switching,
{{\it e.g.}\ } from F to O1 pulsation; this can be, but need not be, the thermal timescale
given by the inverse growth-rate (BK2002).  The longest timescale is
the stellar evolution time, $\tau_{evol}$, {{\it i.e.}\ } the timescale 
corresponding to nuclear burning in the core.

The AE formalism (Buchler \& Goupil \cite{BuchlerG}) was developed to
enable us to study the pulsation problem on the longer timescales, by
averaging out the dynamic timescale, {{\it i.e.}\ } the rapidly varying phases ($\exp i
\omega_k t )$.  The AEs are differential equations which no longer contain the
frequencies, and which govern the behavior of the amplitudes of the most
relevant modes on both the transient and on the evolutionary timescales.

In RR~Lyr the F and the O1 are the only relevant modes
for the pulsation problem.  Furthermore there are no resonances between these
modes, nor with their overtones, and the apposite AEs are (Buchler \& Goupil
\cite{BuchlerG}):
\begin{eqnarray}
\dot A_{\rm 0} &=& \big(\kappa_0(\xi ) + q_{00}(\xi ) A_{\rm 0}^2
               + q_{01}(\xi ) A_{\rm 1}^2 \nonumber \\
	      &&\quad\quad\quad + s_{0}(\xi ) A_{\rm 0}^2 A_{\rm 1}^2 
              + r_{0}(\xi ) A_{\rm 0}^4
                         \big) {\thinspace}\ A_{\rm 0}  
\label{eq_aes1} \\
\dot A_{\rm 1} &=& \big(\kappa_1(\xi ) + q_{10}(\xi ) A_{\rm 0}^2
              + q_{11}(\xi ) A_{\rm 1}^2 \nonumber \\
	      &&\quad\quad\quad + s_{1}(\xi ) A_{\rm 0}^2 A_{\rm 1}^2
              + r_{1}(\xi ) A_{\rm 1}^4
                         \big) {\thinspace}\ A_{\rm 1}
\label{eq_aes2}
\end{eqnarray}
The subscripts 0 and 1 refer to the F and the O1 modes. The quantities
$\kappa_k$ represent the linear growth-rates of the two modes, the $q$s the
cubic, and the $s$'s and $r$'s the quintic coupling coefficients for a given
model. In Koll\'ath~{\it et~al.} (\cite{kollath2}, KBSC) it has been
demonstrated that
we can drop the $r$ terms without prejudice.  
All these quantities are
functions of the model parameters $\xi = (L, {T_{\rm ef\!f} }, M, {\rm
composition})$, where {$T_{\rm ef\!f}$}\ represents the effective temperature of the
equilibrium model.

As far as pulsation modelling is concerned the evolutionary tracks for a given
mass $M$, provide us with a luminosity $L(t)$, and effective temperature
{$T_{\rm ef\!f}(t)$}, i.e. they specify the time-dependence of the $\xi$.  The
stellar envelopes have a uniform composition having recently undergone a fully
convective stage.  We recall that for a given composition, specified by $X$ and
$Z$, for example, these three quantities $M$, $L$ and $T_{\rm ef\!f}$ {\sl
uniquely specify the stellar envelope}, i.e. without a need to know the
properties of the stellar interior.  (One could equivalently, but much less
conveniently characterize the stellar envelope instead by the luminosity,
temperature and pressure at the core radius, for example).  Controversies
associated with core convection and overshooting thus need not concern us here
because pulsation is limited to the part of the star (the envelope) that is
located above the burning shells (i.e. $T<$ a few million K).

%******************************

\begin{figure*}
\centering
\includegraphics[width=8.5cm]{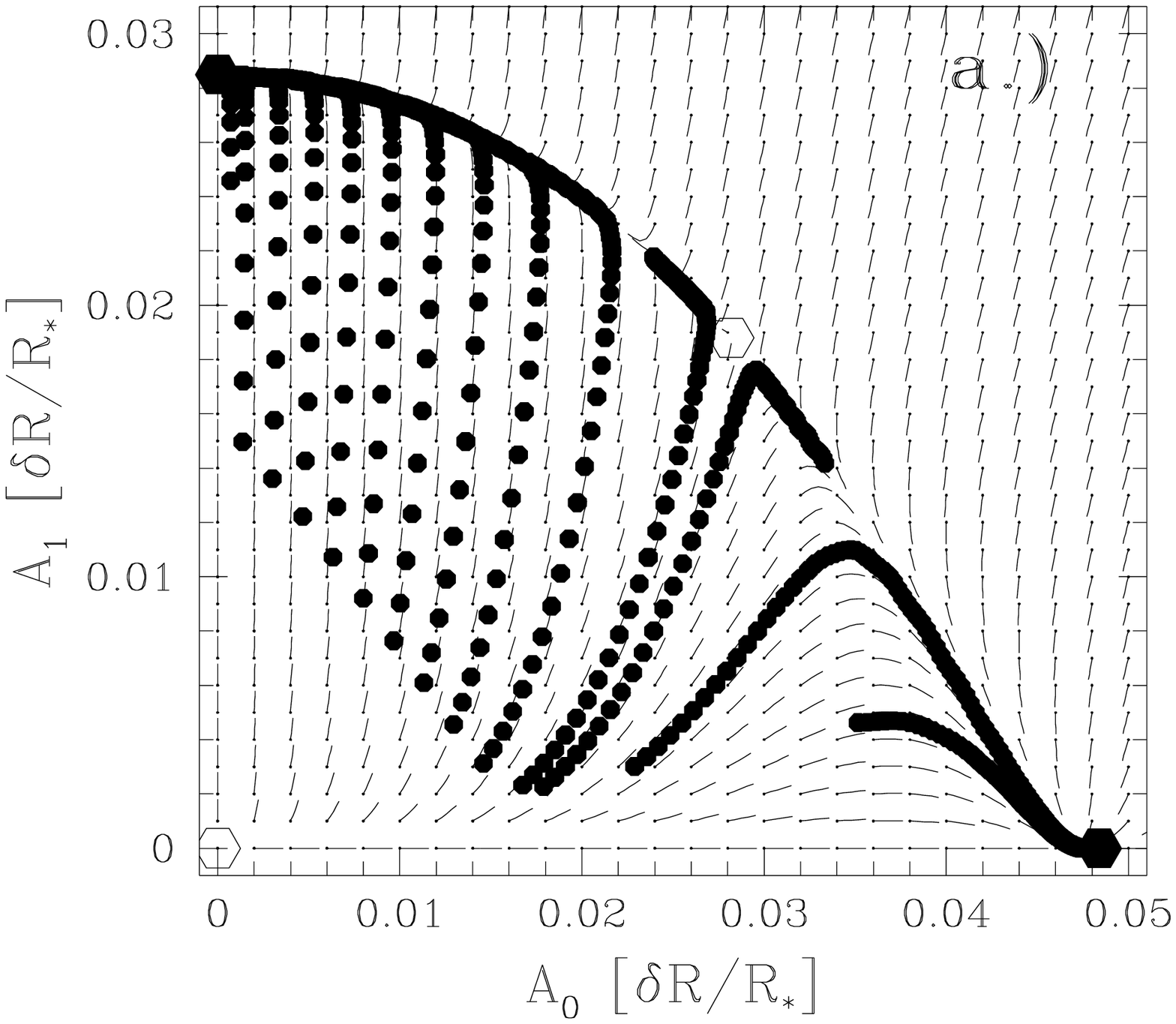}\includegraphics[width=8.5cm]{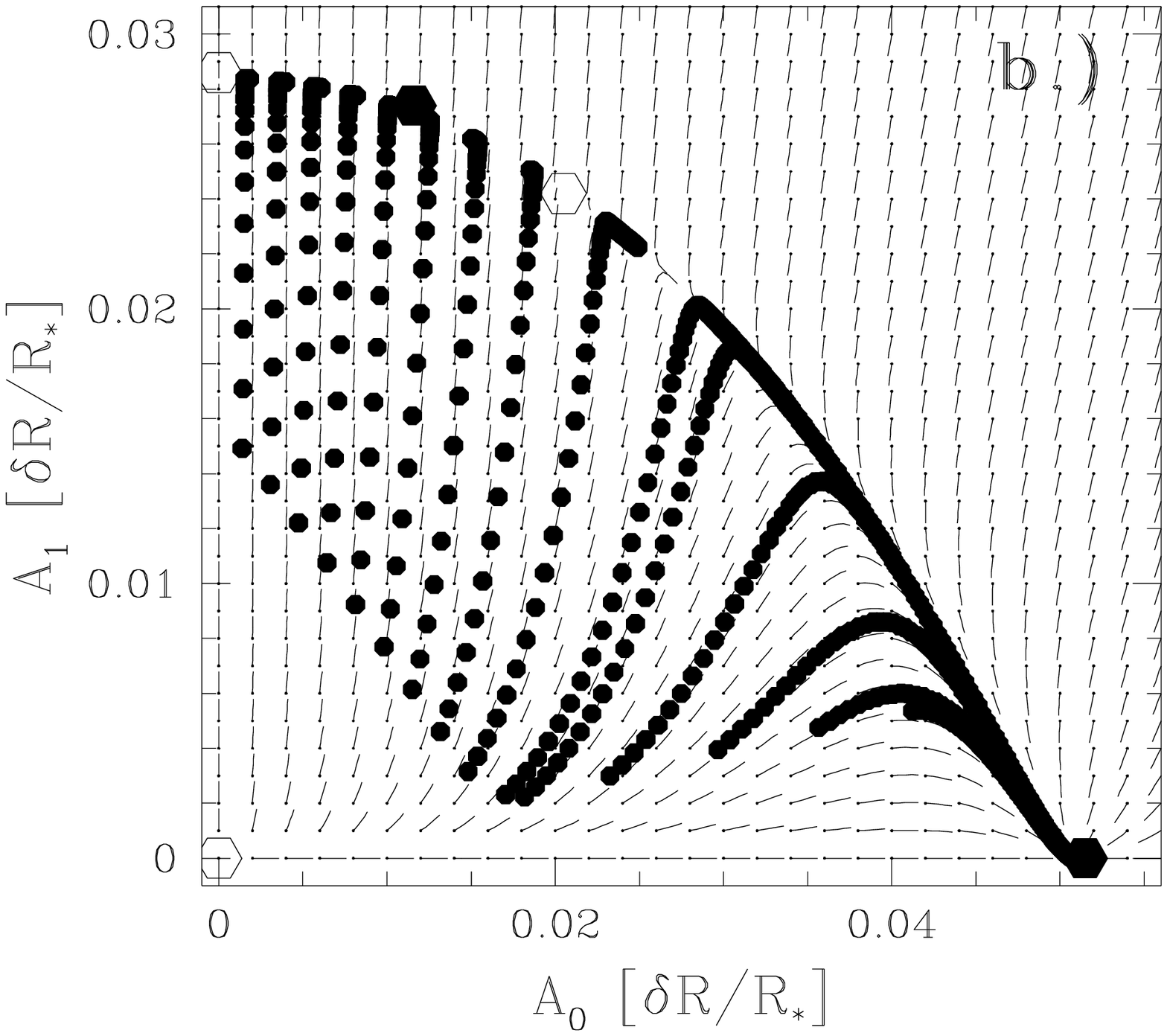}

\caption{Flow vector fields for two RR~Lyr models on the ($A_0,A_1$) phase-space
(the dots denote the bases of the normalized vectors). The large dots
represent the hydrodynamical transients for several different initializations
(kicks). The dots represent equal time-intervals, so that the spacing of the
dots indicates of the speed. {\bf a.)} {\bf F/O1} model ($M=0.71\thinspace {\rm M_{\sun}}$, $L=50\thinspace 
{\rm L_{\sun}}$, $T_{\rm ef\!f} = 6700 {{\thinspace} {\rm K}}$, $Z=0.0001$). The stable F and O1 limit cycles 
denoted by filled hexagons are located at ($0.0483 \thinspace ;0.0000$) and ($0.0000 \thinspace ;0.0285$), 
respectively, and the presence of an unstable DM (open hexagon) can be seen on the arc at 
($0.0282 \thinspace ;0.0188$). {\bf b.)} {\bf F/DM} model ($M=0.77\thinspace {\rm M_{\sun}}$, $L=50\thinspace 
{\rm L_{\sun}}$, $T_{\rm ef\!f} = 6510 {{\thinspace} {\rm K}}$, $Z=0.001$). The stable fundamental and 
double-mode fixed points are located at ($0.0515 \thinspace ;0.0000$) and ($0.0116 \thinspace
;0.0274$), respectively. The unstable O1 and DM fixed points can be found at 
($0.0000 \thinspace ;0.0286$) and ($0.0205 \thinspace ;0.0242$).}

\label{fig1}
\end{figure*}

%******************************

It is useful to introduce the quasi-static approximation (QSA) which consists
of disregarding the time-dependence of the $\xi$.  The stellar model is thus
considered frozen at a point along its evolutionary track.  The QSA serves two
purposes.  First, the AEs with the QSA,
i.e. Eqs.~\ref{eq_aes1}--~\ref{eq_aes2}. with fixed constant coefficients,
describe the temporal (transient) behavior of the modal amplitudes of a stellar
model {\sl with fixed} $M$, $L$ and {$T_{\rm ef\!f}$}.  The QSA thus provides
an excellent description of the transient pulsational behavior of a given
stellar model from some initial conditions toward its asymptotic limit cycle or
DM pulsation, as the case may be.  Conversely, the hydrodynamic evolution of
such a model can be employed to extract the coefficients of the AEs for given $M$,
$L$ and {$T_{\rm ef\!f}$}.

Second, away from the bifurcation points, $\tau_{trans} \ll \tau_{evol}$, and the
stellar model should be in (one of) its asymptotic pulsational states.
Instead of having to integrate Eqs.~\ref{eq_aes1}--~\ref{eq_aes2}. one then only
needs to calculate their fixed points, $\dot A_k=0$, which correspond either to
limit cycles or to steady double-mode pulsations.
When more than one asymptotic pulsational state
is possible the star naturally stays in the one that is closest to the one it
had at an earlier instant.  Hysteresis is possible when evolutionary tracks run
in opposite directions at different stages of the evolution.

In summary then, if we can compute the $\kappa_k$ and the coupling coefficients
as a function of $L$ and {$T_{\rm ef\!f}$}\ from pulsation theory on the one
hand, and the time-dependence of the latter from the evolutionary tracks, on
the other, we can then integrate Eqs.~\ref{eq_aes1}--~\ref{eq_aes2}. to obtain
$A_k(t)$ along the whole track, including mode switchings, also called
bifurcations (BK2002).

We note in passing that the more complicated situation $\tau_{trans} \sim
\tau_{evol}$ that always occurs near changes of stability (bifurcation points)
has recently been examined by BK2002.

% =============================================

\subsection{The Nonlinear Coupling Coefficients}

One finds that the growth-rates and the coefficients vary smoothly with
{$T_{\rm ef\!f}$}, $L$, and $M$.  Consequently it is sufficient to compute these
coefficients for only a few stellar models with fixed {$T_{\rm ef\!f}$}\ and $L$ inside the
IS, and the intermediate values along the track can then be
obtained by interpolation.  In sharp contrast, the amplitudes,
solutions of the AEs, and the pulsational states can vary very rapidly with
{$T_{\rm ef\!f}$}\ and $L$.

The computation of the coupling coefficients of a given model, with a specified
{$T_{\rm ef\!f} $}, $L$ and composition proceeds in a few steps (Koll\'ath \&
Buchler \cite{kb01}).

% =======================================

\subsubsection{The hydrodynamical calculations}

The star can reach more than one type of full amplitude pulsation depending on
the initial conditions (several basins of attraction).  The
most familiar case is the coexistence of either F or O1 limit cycles, but they can
also be steady double-mode (DM) pulsations.  In that case it is important to
choose initial perturbations so that the transient tracks sample 
enough of the amplitude-amplitude plane.
Thus, for several suitably chosen initial perturbations of the (pulsationally
unstable) equilibrium model we follow its transient pulsational behavior with
numerical hydrodynamics until the achievement of steady, or almost steady full
amplitude pulsations.

In this paper we investigate pulsation from dynamical point of view, and we do
not consider light-curves and velocity curves. The interested reader is
referred to Koll\'ath \& Buchler (\cite{kb01}). We note that light and velocity
curves are in good agreement with the observations and an accurate radiative transfer 
rather than radiation diffusion as we use has little effect on the bolometric light
curves. More detailed computations from the point of view of color dependence of LCs
have been made by Dorfi \& Feuchtinger (\cite{df}).

% =======================================

\subsubsection{The analytical signal method}

The output from each of the hydrodynamical calculations, for example the
stellar radius $R_*(t)$, is subjected to the analytical signal (time-frequency)
analysis (G\'abor \cite{gabor}).  This provides the
time-dependence of the amplitudes $A_k(t)$ (KBSC).  Astrophysicists are generally 
familiar with this procedure
because of the Kramers--Kronig dispersion relations (Jackson \cite{jackson}).
One considers the signal $s(t)$ to represent the real part of an assumed
complex {\it analytical} function $a(t)$.  The imaginary part $\tilde s(t)$ of
$a(t)$ can then be obtained {\it via\ } Cauchy's formula, which is
converted to a Hilbert transform, and finally to a Fourier transform and a subsequent
one-sided reverse Fourier transform.  The advantage of the latter formulation is that both are
extremely fast to perform ({\it cf.} {{\it e.g.}\ }Cohen \cite{cohen}).
%************************************** 
\begin{eqnarray} 
a(t) &
\negthinspace\negthinspace\negthinspace =
\negthinspace\negthinspace\negthinspace &s(t)+i \tilde s(t) 
   = s(t)+\frac{i}{\pi} {\thinspace} PV{\thinspace} \int_{-\infty}^\infty dt' 
   \frac{s(t')}{t-t'} \nonumber\\ 
 & \negthinspace\negthinspace\negthinspace 
= \negthinspace\negthinspace\negthinspace & \frac{1}{\pi } 
  \int^{\infty}_{0}d\omega e^{i\omega t} \int_{-\infty}^\infty dt' 
    s(t')e^{-i\omega t'} \nonumber\\ 
&\negthinspace\negthinspace\negthinspace 
 \equiv \negthinspace\negthinspace\negthinspace 
 & A(t){\thinspace}{\thinspace} e^{i\varphi (t)} 
\end{eqnarray} 
%************************************** 
With this procedure it is therefore very easy to determine unambiguously the
instantaneous phase $\varphi(t)$ and amplitude $A(t)$ of a signal.

The method is extended to multi-component signals with the help of filters
that restrict the power to the desired frequency components.  It is convenient
to make this filtering in Fourier space where it involves just a product, and
to combine it with the definition of the analytical signal
%**************************
\begin{eqnarray}
Z_{k}(t) & \negthinspace\negthinspace\negthinspace =
\negthinspace\negthinspace\negthinspace 
           & A_{k}(t)e^{i\varphi_{k}(t)} \nonumber \\
           & \negthinspace\negthinspace\negthinspace =
	   \negthinspace\negthinspace\negthinspace 
           & \frac{1}{\pi }\int ^{\infty
	   }_{0} \negthinspace\negthinspace\negthinspace d\omega
           H(\omega -\omega _{k})e^{i\omega t}
           \negthinspace\int_{-\infty}^\infty
	   \negthinspace\negthinspace\negthinspace dt' 
              s(t')e^{-i\omega t'},
 \label{eqaf}
\end{eqnarray}
%**************************
where $H(\omega -\omega _{k}) $ is the filtering window, centered on the
desired $\omega_k$.  Generally, a Gaussian window with a half width of 0.2 c/d
provides satisfactory results.

{\bf Figure}~\ref{fig1}a. illustrates the amplitude evolution of an RR~Lyr model
that has both an F and an O1 limit cycle.  The origin is unstable (equilibrium
model), and depending on the initial kick the model goes to one or the other
limit cycle.  In addition to these limit cycles, there exists an unstable DM
that is clearly visible along the arc. An example of another, more complicated
phase-portrait involving two DMs can be found in {\bf Fig.}~1b. There 
are two stable fixed points here: a DM and an F. 
This stable DM pulsation exhibits a dominant overtone pulsation amplitude. 

%******************************
\begin{figure*}
\includegraphics[height=7cm,width=16.0cm]{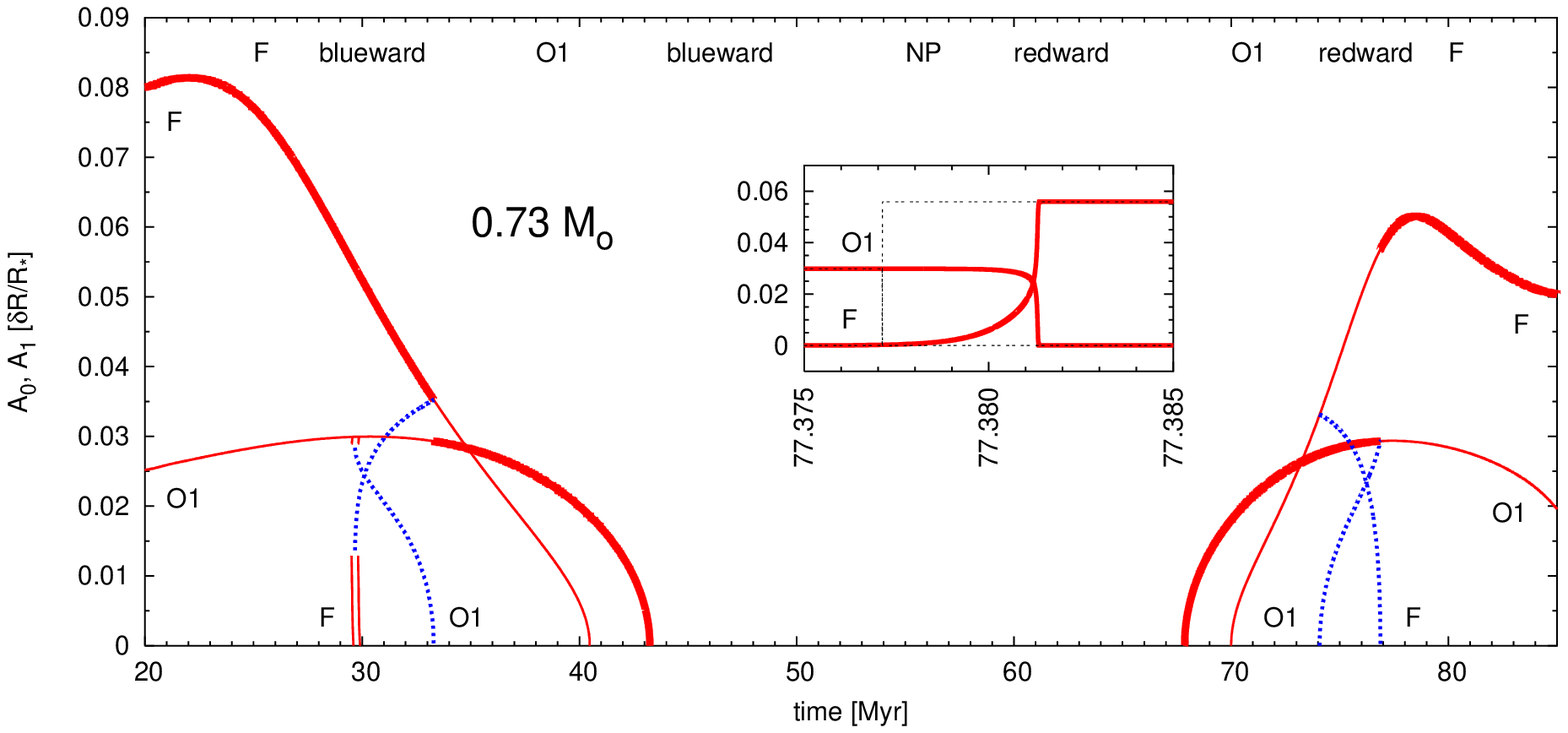}\\
\includegraphics[height=7cm,width=16.0cm]{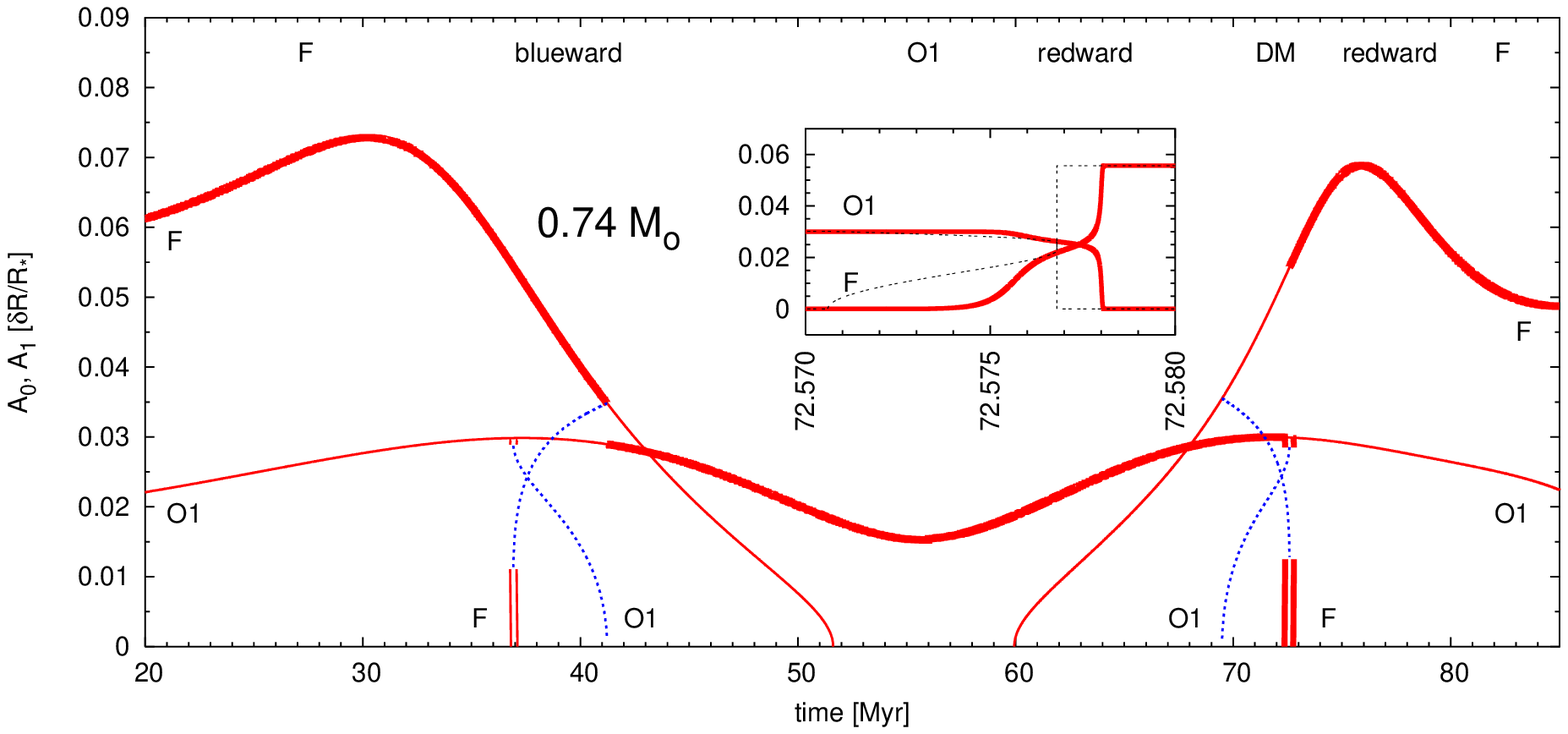}
\caption{Evolution of the amplitudes and component amplitudes of two RR~Lyr
models (with $Z=0.0001$) throughout the IS; {\sl upper panel:}
$M=0.73\thinspace {\rm M_{\sun}}$ {\sl lower panel:} $M=0.74\thinspace {\rm M_{\sun}}$.  
The thick lines represent
the pulsational modal amplitudes that actually achieved attained along the
track, the thin lines potential pulsation states, the dotted lines the
amplitudes of {\sl unstable} double-mode pulsation.  The stable DM amplitudes
are represented by double lines.  DM pulsation 
occurs during the redward evolution of the higher mass model.  The attained
pulsation mode and the direction of the evolution are indicated on top; NP
denotes the non-pulsating interval during which the star leaves the instability
strip.  The insets represent blowups of the amplitudes during the redward mode
switching phases that last a few thousand years. The broken lines in the insets
show the QSA amplitudes, i.e. the amplitudes that would occur if the pulsation
could adjust instantaneously to the asymptotic limit cycle of DM pulsation 
of the model with the current $M$, $L$ and $T_{\rm ef\!f}$.  
Compare with {\bf Fig.}~\ref{fig5}.}
\label{fig2}
\end{figure*}
%******************************

% =======================================

\subsubsection{The linear least-squares fit}

Once the temporal behavior of the amplitudes has been obtained we can determine
the coupling coefficients and the linear growth-rates with a linear
least-squares fit from Eqs.~\ref{eq_aes1}--~\ref{eq_aes2}. in which $\xi$ is
fixed to the values of $M$, $L$ and {$T_{\rm ef\!f}$}\ that have been used in the
hydrodynamical computations.

The fit is linear because the analytical signal analysis yields amplitudes that
are smooth enough to compute time-derivatives, and it is not necessary to fit
the transients to integral curves of Eqs.~\ref{eq_aes1}--~\ref{eq_aes2}.  as
with the older technique (Buchler \& Kov\'acs \cite{bk87}; Kov\'acs \& Buchler
\cite{bk93}; Kov\'acs~{\it et~al.}  \cite{kbd}).

% =======================================

\subsubsection{The evolution of the pulsations}

The previous  
procedure can be repeated for a sequence of models along the evolutionary
track and the behavior of the growth-rates and coupling coefficients as a
function of {$T_{\rm ef\!f} $} and $L$ can then be obtained by interpolation.
The solution of Eqs.~\ref{eq_aes1}--~\ref{eq_aes2}. allows us to follow the
pulsational behavior of our model along its evolutionary track once we know the
temporal behavior of the $\kappa$s and of the coefficients because of slow
stellar evolution.  For that transformation we need $T_{\rm ef\!f}(t)$ and
$L(t)$ which are obtained from published stellar evolution calculations, {{\it
e.g.}\ } Schaller~{\it et~al.} (\cite{ssmm}); Alibert~{\it et~al.}
(\cite{alibert}) for Cepheids, Dorman (\cite{dorman}); Demarque~{\it et~al.}
(\cite{demarque}); Girardi~{\it et~al.} (\cite{girardi}) for RR~Lyr.

In {\bf Fig.}~\ref{fig2}. we present the evolution of the pulsation
amplitudes of two RR~Lyr models along their Demarque evolution tracks
through the IS which are shown in {\bf Fig.}~\ref{fig5}.  The
$M=0.73\thinspace 
{\rm M_{\sun}}$ model starts up with F mode pulsation.  Then, while evolving blueward,
it changes to O1, and it finally leaves the IS temporarily.
Next, it
repeats these steps in reverse order while evolving redward.  The higher mass
model also experiences DM pulsation for a short period during its redward
evolution, entering the F/DM hysteresis region (Sec.~3.2). We note that we
have not made the QSA for the amplitudes and that we have considered the
evolution induced delay in the switching (BK2002).

The insets of {\bf Fig.}~\ref{fig2}. enlarge the interval over which the righthand
switchings take place.  A scenario similar to the top inset, but with O1 and F
interchanged takes place near 33\thinspace Myr for the $0.73\thinspace {\rm M_{\sun}}$ 
model, and near 41\thinspace Myr for the $0.74\thinspace {\rm M_{\sun}}$ model. 
For all of these switchings, which we termed {\sl
hard} in BK2002, the timescale is of the order of a few thousand years, much
shorter than the evolution timescale, and somewhat longer than the thermal
timescale.  The broken lines in the insets present the QSA amplitudes, {{\it i.e.}\ } the
amplitudes that would obtain if the switch was instantaneous.

In the top inset the switch occurs from a stable O1 mode limit cycle to
a stable F mode limit cycle. The transition is not sharp and there is a delay,
both caused by evolution as explained in BK2002.

The situation in the bottom inset is more complicated.  Here the switch
occurs first from an O1 limit cycle to a DM state, then from the DM to the F
limit cycle.  Again, the two transitions are not sharp, but washed out by
an  evolution induced delay in the switching.
The DM pulsation is in principle already possible at $\sim$72.5705 Myr, but
because of evolution it gets delayed  $\sim$4000 yrs.  Subsequently, the F
pulsation could start at $\sim$72.577, but gets delayed some 1000 years.

Observationally it would be very hard to distinguish between the switching from
RRab to RRc (or vice-versa, as displayed in the inset in the top figure) on the one
hand, and
true RRd behavior, as seen in the bottom inset, on the other hand, 
although from a theoretical
point of view they are very different.

{\bf Figure}~\ref{fig2}. clearly demonstrates that the time spent in the mode
switching phases is very short compared to the RRab and RRc phases.

% =======================================

\subsubsection{Resonances}

It is well established that as the structure of the stellar envelope changes
along the evolutionary track, resonances can occur between the excited mode and
one or more overtones.  The presence of resonances has a strong effect on the
Fourier decomposition parameters of the light-curves and radial velocity curves
({{\it e.g.}\ } Buchler \cite{Mito}).  In this paper we ignore the effects of
resonances which is a valid assumption for RR~Lyr stars, but would not be for
Cepheids.

\vskip 10pt 

In summary, our new methodology makes it  
possible to follow in an automated fashion the full
amplitude pulsational behavior along the tracks that stellar evolution
computations provide.

%======================================================================

\section{Nonlinear RR~Lyrae Models}
\label{sec3}

% =============================================

\subsection{Turbulent convective models}
\label{subsec3.1}

In order to explore the mode selection characteristics in numerical RR~Lyr
models, we have used the Florida-Budapest TC-codes ({{\it e.g.}\ } Yecko~{\it et~al.} 
\cite{ykb}, KBSC), which are 1D, linear and nonlinear hydrocodes 
that include turbulent convection and are tailored to radial stellar pulsations.

The hydrogen content of the models is set to be $X=0.75$.  The turbulent
convection parameters are chosen as in KBSC (Table~1.) to ensure observed
amplitudes. Model sequences for each combination of triplets 
($L$, $M$, $Z$) from
%**************************
\begin{itemize}
\item{$M = 0.50, 0.55, 0.60, 0.65, 0.71, 0.77, 0.82, (0.85), 0.87 {\rm M_{\sun}}$}
\item{$L = 40, \ 50, \ 60, \ 70 \ {\rm L_{\sun}}$}
\item{$Z = 10^{-4}, \ 10^{-3}, \ 4\cdot 10^{-3}$}
\end{itemize}
%**************************
have been computed, in which only {$T_{\rm ef\!f}$}\ is varied. No evolutionary constraints 
were applied in the parameter selection at this point. Applying the 
methodology described in Sec.~{\ref{sec2}}. has allowed the interpolation of 
the limits of different pulsational states to an internal accuracy of $2-3 {{\thinspace}
{\rm K}}$ within a sequence, requiring the computation of only 8 to 10 models in a given
sequence. 

We have generated a large grid of models which contains mode selection
information throughout the relevant regions of the parameter space. We note here again, that
at this point we do not care about the existence of the objects represented by these 
models, as the whole model set is used only to point out mode selection characteristics (Sec.~4.). 
The existence will be ensured in Secs.~5-6., where evolutionary tracks are applied to select the 
astronomically relevant models among the complete set of computed models. 
This way, the presented results concerning the 
fundamental blue edge and the double-mode RR~Lyrae stars are based on pulsational models that 
were naturally selected by evolutionary calculation constraints. 

The following combinations of one or two stable fixed points are encountered in our models:

%**************************
\begin{itemize}
\item{fundamental mode (F)}
\item{first overtone (O1)}
\item{either first overtone or fundamental mode  (F/O1)}
\item{double-mode (DM)}
\item{either fundamental or double-mode (F/DM)}
\end{itemize}
%**************************

Because the DM and F/DM regions are very narrow, they are difficult to find
with hydrodynamical modelling, although Feuchtinger (\cite{feuchtinger1})
had the good fortune to actually encounter one!  
However, interpolation on our grid easily reveals
their presence.  Once their existence and their astrophysical parameters are
known, it is not difficult to confirm the presence of DM pulsation through the
computation of further models.  We conclude that amplitude equation fitting is a
robust and effective method in this sense.

We now turn to a discussion of the shortcomings and uncertainties in our model
computations. The selection of turbulent convective parameters is not unique.
We have found that the amplitude and the IS width requirements cannot be
fulfilled at the same time for a given set of $\alpha$ parameters. We have
therefore fixed turbulent parameters to give the correct observed amplitudes
(magnitudes) as in KBSC (Table~1.), but that extended the location of the F
linear red edge to very low temperatures, giving an excessive width to the F
IS. On the other hand, it caused but a small excursion in all the other
edges. If we were to apply a different set of $\alpha$s, a fine-tuning of the
TC parameters would be necessary, and that would bring about small temperature shifts
of $10 - 50 {{\thinspace} {\rm K}}$. But these shifts would affect the mode 
selection regions as a whole, not changing the location of the edges with 
respect to each other (except F red edge). Therefore we think that, aside from 
small shifts in temperature our overall picture is correct, and we attained a 
good relative accuracy in determining mode selection.

% =============================================

\subsection{Mode selection hysteresis}

%**************************

\begin{figure*}
\includegraphics[width=14cm]{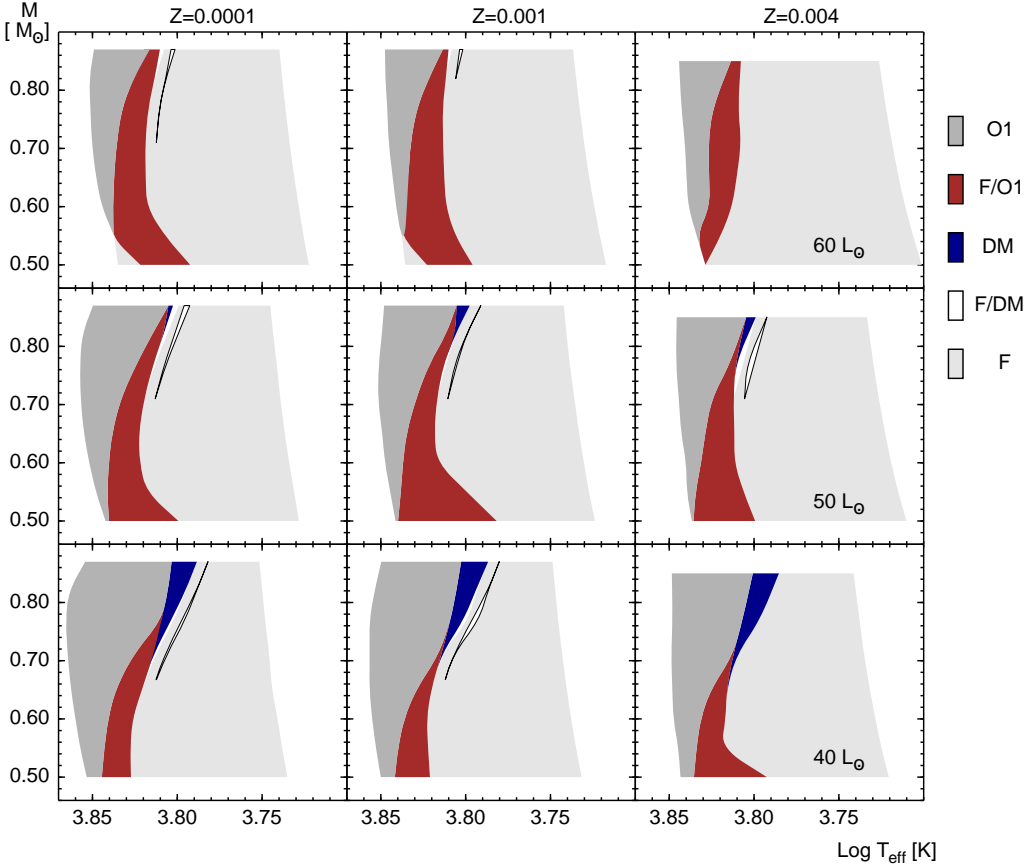}
\caption{ \ Overview of the pulsation modes for RR~Lyr stars as a function of 
stellar mass. Rows have fixed luminosity, columns contain fixed metallicity. 
Note that the models displayed  in this figure were computed with arbitrary selections 
of mass, luminosity, temperature and metallicity to fully explore the mode selection 
characteristics. See text for details. The legend is the following: 
F: fundamental mode 
F/DM: either fundamental or double-mode 
DM: double-mode 
F/O1: either fundamental or first overtone (F/O1) 
O1: first overtone mode.  
\break {\bf Note that for the sake of clarity 
the F/DM region has been plotted shifted by 100 ${{\thinspace} {\rm K}}$ to the right 
in the Figures.} }
\label{fig3}
\end{figure*}
%**************************

It is well known from the early suggestion of van Albada \& Baker
(\cite{albada}) that in the case of classical pulsators there exists a region
in the HR diagram in which either F or O1 pulsations are possible.  Rather than
follow the pulsation jargon of calling it the 'either-or-region', we refer to
it as a F/O1 region, especially since we will also encounter other types of
hysteresis.  Hysteresis here means that the pulsation state depends on the
direction of evolution.  For example,  if the star evolves from low temperatures 
to high temperatures (blueward) in this region, it exhibits F mode pulsation 
(RRab in our case), while evolving in the opposite direction results in an O1 
pulsator (RRc) with the same stellar parameters.  This is the classical F/O1 
region. Recent observational facts (Castellani~{\it et~al.}  \cite{castellani}), 
the amplitude equation formalism (Buchler \& Kov\'acs \cite{bk}) as well as
state-of-the-art numerical hydrodynamical calculations (KBSC) all concur that 
this behavior indeed exists.

It is interesting to note that we have encountered an additional hysteresis in
the course of the present survey, namely an F/DM region where \emph{ either F or
DM} pulsations are possible.

The width of the F/O1 region can be several hundred \thinspace ${\rm K}$, depending
on the mass, luminosity, and metallicity. As stars with different parameters
pass through this region, mode selection can significantly bend the
\emph{average slope} of the observed F blue edge. Therefore this effect
supplemented by reliable evolutionary data has the potential of reproducing the
slope of the RR~Lyr F blue edge. One of the main motivations of this paper
(Sec.~\ref{sec5}) has
been to investigate this scenario as a possible resolution of the puzzle that
was discussed in KBF.

%======================================================================

\section{Mode selection maps}
\label{sec4}

The computed mode selection maps are exhibited  in {\bf Fig.}~\ref{fig3}.,
presented in the form of a $M$ -- {$T_{\rm ef\!f}$}\ plot. For the first time, double-mode regions 
are also included. We emphasize that the models of  {\bf Fig.}~\ref{fig3}. were computed with 
arbitrary selections of mass, luminosity, temperature and metallicity, as was described in 
Sec.~{\ref{subsec3.1}}. Although some configurations of $M$, $L$, and $T_{\rm ef\!f}$
may well not actually exist in Nature, 
this method 
allows us to explore the mode selection characteristics throughout the whole parameters space. In 
addition, by reducing the parameter space (e.g. by applying evolutionary constraints) in order to 
minimize the time-consuming model computations, we could easily miss important features of the mode 
selection phenomenon. These models are later subject to merge with evolutionary tracks to end up 
with models of existing objects (Secs.~\ref{sec5}-\ref{sec6}.) in a self-consistent way. We now 
discuss the most important mode selection features:
%**************************

\vskip 5pt
 \noindent -- The topology of the IS is very similar but not identical
for different metallicity values. Clear differences also exist in different
luminosity panels and for different masses. It is in good agreement with the results of Bono
\& Stellingwerf (\cite{bono94}), namely with a redder F blue edge at lower
luminosity levels.  In terms of metallicity the difference manifests itself in
the form of slight shifts in temperature, but typically less than $100
{{\thinspace} {\rm K}}$.  We caution though, that these small changes cannot
be neglected when we derive the slope of the IS edges, as 
will be pointed out in Sec.~{\ref{sec5}}.
 
 \noindent -- The O1 blue edges are consistent with those of Catelan (\cite{catelan})
and references therein, but depend on mass and metallicity and to a lesser
extent on luminosity.
 
 \noindent -- F/O1 behavior is found across the whole range of our $Z$ and $L$ values.
This region is of primary interest in blue edge investigations. We note the
higher uncertainty of mode selection in the low mass region, and the wider F/O1
regions, for example, have to be taken with some caution.

 \noindent -- A double-mode region (DMR, by which we denote both the DM and F/DM
regions) is also present throughout all panels, generally with a very small
width: $10-60 \thinspace \mathrm{K}$ at maximum.  Although it is missing from the upper
right panel ($Z=0.004$, $L=60\thinspace {\rm L_{\sun}}$), one could obviously find 
it for even higher masses.  For $L=40\thinspace {\rm L_{\sun}}$ we obtain mostly a DM 
only area, while at higher luminosity a F/DM region appears immediately at the lower 
temperature side of the DM domain.

Note that in {\bf Fig.}~\ref{fig3}. the F/DM regions are extremely narrow and hug the
pure DM regions; for clarity they are plotted shifted by $100 {{\thinspace} {\rm K}}$
to the right.

 \noindent -- The positions of DMR and F/O1 are also
tightly related in that  they always touch
and are situated between pure F and O1 regions.  For fixed $L$ and $Z$
a F/O1 region exists at low mass and a DMR at higher mass.  It is interesting
how a DMR moves into the place of the F/O1 region at low luminosities.  The
transition mass ({{\it i.e.}} minimum DM mass) increases with increasing
luminosity.
 
 \noindent -- A gradual narrowing of the pure O1 region is found toward low masses.
On the other hand the net area of possible O1 pulsation (O1 + F/O1) remains
more or less constant.
 
 \noindent -- At high $L$ and low $M$, an additional F region can be seen, that is
equally present on all $L=60\thinspace {\rm L_{\sun}}$ panels.  The O1 and the extra 
F regions are separated by the F/O1 region ({{\it i.e.}\ } there is a triple point 
in the ${T_{\rm ef\!f} } - M$ plane).  While this is allowed by the
amplitude equations, it does not play a role in determining the slope of the
blue edge, as no evolutionary tracks reach this region. 
 
 \noindent We have not encountered O2 limit cycles (RRe stars) in our survey, which
concentrated on F, O1 and DM pulsations.  A specific study of O2 pulsations
will be taken up in future work.

%======================================================================

\section{Improved theoretical fundamental blue edge}
\label{sec5}

With our methodology we have been able to generate detailed mode selection maps,
including even the hard to find DMRs.  However, in order to gain insight into the
interplay of pulsation and evolution it is necessary to include evolutionary
effects, and to perform a population synthesis.  

\subsection{Synthetic instability strip}

Thus we construct a \emph{synthetic} 
${\rm Log}\thinspace {T_{\rm ef\!f} } - {\rm Log}\thinspace L$ plot, in which we 
combine the HB evolutionary sequences and account for hysteresis in order to derive 
a slope for the F blue edge that can be compared to the observational data. It is 
worth mentioning that synthetic horizontal branches have been made earlier 
({{\it e.g.}\ } Lee {\it et~al.} \cite{lee}).
Bono~{\it et~al.} (\cite{bono97}) computed nonlinear RR~Lyr pulsational 
models for three representative mass values and five different luminosity levels accounting 
for the evolution. Their method is useful for modelling globular cluster ISs
for a given population. In order to model a 'composite' blue edge and to
compare it to an empirical one (Jurcsik \cite{jurcsik}) representing the blue 
edge of a mixed population of RR~Lyrae stars (field, LMC, globular cluster
etc.), we combine these two approaches here, but on a broad scale.  The large
number of our models enables us to interpolate between the computed mass,
luminosity and metallicity values, and we get a continuous and smooth distribution of stars 
in the IS. In this way a self-consistent, simultaneous treatment of 
the mode selection and evolution is possible, and even the theoretical definition of the 
fundamental blue edge is made more consistent. 

Three HB evolutionary tracks are compared in order to uncover common 
trends and conclusions:
\begin{itemize} 
 \item{oxygen-rich Dorman-tracks (Dorman \cite{dorman}),} 
 \item{Demarque-tracks (Demarque {\it et~al.} \cite{demarque}) and} 
 \item{Padova-tracks (Girardi {\it et~al.}
\cite{girardi}).}  
\end{itemize} 
Only slight differences exist between the Dorman and Demarque tracks.  For a
detailed comparison of Demarque and Padova evolutionary tracks the reader is
referred to a recent paper (Gallart~{\it et~al.} \cite{gallart}).  Evolutionary
tracks are interpolated in age here to generate a high-resolution,
homogeneous dataset in each case.

Combining the mode selection and evolutionary information we construct a
theoretical ${T_{\rm ef\!f} }-L$ diagram for RR~Lyr stars pulsating in the F
mode and in the O1 mode. To this end a synthetic horizontal branch stellar
population is generated with arbitrary mass, metallicity and age distributions.
For each individual model star, according to its parameters ($M,Z;t$), we then 
determine its evolutionary track and the corresponding pulsational properties by  
interpolation in the evolutionary and pulsational grids, respectively. This way the 
independently computed pulsational models and evolutionary results are matched in a 
consistent way (the consistency is discussed in Sec.~\ref{msc}.), and from now on 
(as opposed to Sec.~\ref{sec4}.) the pulsational characteristics are governed by and 
chosen to match the evolutionary criteria.

It is found convenient to
determine the temperature boundaries of the various pulsational regimes (blue
and red edges of O1, F/O1, DM, F/DM, F) through interpolation.  The pulsational
states corresponding to every model star are then determined by taking into account
the interpolated widths of the F/O1 region.  Outside the F/O1 regions there is
no ambiguity and every point is considered an F or O1 pulsator depending on its
effective temperature.  Inside, where there are two possible pulsational states,
the direction of evolution decides which one the model pulsates in.

The census in this section 
omits the DM stars, first, because the pulsational state
after the traversal of a DMR is the same as it would be as if there were no
DMR, and second, the DMRs are very narrow and hardly affect the proportion of F
and O1 stars.

At first glance the time delay for switching from one pulsational state to
another should also be taken into account when deriving the slope of the F blue
edge (BK2002), because when entering the IS the star does not immediately
achieve pulsations with the amplitudes of the pulsational mode that are
expected in the QSA.  Similarly, when crossing the borders between pulsational
modes (bifurcations), there is a delay in switching to the new pulsational
state as we have seen in {\bf Fig.}~\ref{fig2}.
Therefore, a small number of stars could be missing at the boundaries of
the instability zone, or be slightly shifted.  However, the timescale of the
delay (thousands of years, {\it cf.\ } inset of {\bf Fig.}~\ref{fig2}.)
is much shorter than the crossing time of the F/O1 region (orders of $10^6$ yrs).

As we intend to establish a \emph{universal} F blue edge, in the sense that an
ensemble of RR~Lyr stars of all possible masses, heavy element abundances and
ages are included, we have decided to restrict ourselves to uniform
distributions.  One could argue, that a (truncated) Gaussian distribution is
more appropriate for a stellar mass distribution along the horizontal branches,
but (1) we have very sparse information about the mass of the observed stars
used in KBF, (2) it can be demonstrated
that the mass-distribution
has little effect on the IS boundaries in the case of
\emph{universal} F blue edge.  Only the number density and the distribution of
the synthetic stars change on the 
${\rm Log}\thinspace{T_{\rm ef\!f} } - {\rm Log}\thinspace L$
diagram.  We have performed several tests to verify this conclusion.  Note that
we also have assumed a uniform age distribution (with age measured as time
elapsed since leaving the ZAHB).

\subsection{The Model Selection Criteria}
\label{msc}

%**************************
\begin{figure}
\resizebox{\hsize}{!}{\includegraphics{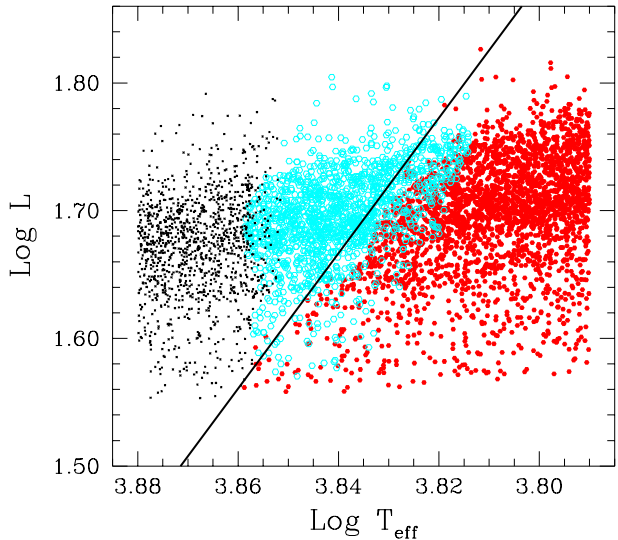}}

\caption{Synthetic IS based on Demarque evolutionary tracks and
our pulsational models. Filled points represent F mode RR~Lyr,
open circles overtone RR~Lyr, and small dots nonpulsating (hot) HB-stars.
The thick line delineates the observed F blue edge.  Notice the remarkable
agreement between the empirical F blue edges and the envelope of the
F pulsators.  See text for details.}

\label{fig4}
\end{figure}
%**************************

Our luminosity range is $1.5 < {\rm Log}\thinspace {\it L} < 1.86$, the mass 
interval spans $0.57 M_{\odot} < M < 0.87 M_{\odot}$, and the metallicity varies 
from $Z=0.0001 - 0.004$, in such a way that the distribution is uniform in the
corresponding $[Fe/H]$ interval.  Extrapolation has not been extended beyond
these limits to reduce uncertainties.  These mass values encompass those
expected for 
RR~Lyr stars.  The age distribution (measured from the ZAHB) has been truncated
at $t= 90.0-95.0$~ Myrs, depending on the characteristics of the evolutionary
track, in order to decrease uncertainties in interpolation, and to avoid the
high luminosity region, where the linear trend breaks down according to our
investigations.  We note that this truncation is compatible with the above
mentioned luminosity maximum.

One example of a synthetic 
${\rm Log}\thinspace {T_{\rm ef\!f} } - {\rm Log}\thinspace L$ plot based on
Demarque's evolutionary tracks and on our turbulent convective, nonlinear
models is displayed in {\bf Fig.}~\ref{fig4}.  Plots based on Dorman and Padova
evolutionary tracks are quite similar. Filled points represent F mode RR~Lyr,
and open circles O1 RR~Lyr. Small dots denote nonpulsating HB-stars.
The effective temperature interval on the plot is chosen to be the same as in
{\bf Fig.}~3. of KBF for comparability. We note that our models and
evolutionary tracks are compatible with this selection. The empirical
F blue edge is indicated by a straight line. It is adapted from Jurcsik
(\cite{jurcsik}). The RR~Lyrae sample is taken from Jurcsik (\cite{jurcsik97}), while 
the empirical relations supplying transformation from the empirical to the theoretical
plane are taken from Kov\'acs \& Jurcsik (\cite{kj1}, \cite{kj2}) and Jurcsik 
(\cite{jurcsik}). The precursor of the present work, KBF (\cite{kbf}) uses the same 
observational sample and transformations, and furthermore discusses 
the quality of the agreement between observations and models, as well as the uncertainties 
related to both the model computations and the empirical relations. 
Because this work closely follows the same method we refer the interested 
reader to the cited  work for further details.  

Out of 45\,000 artificially generated HB-stars, the diagram contains 2679 F,
1455 O1 and 1457 nonpulsating stars.  We have performed 10\,000 Monte-Carlo
iterations for all three evolutionary track sets.  The slopes of the linear
fits are listed in {\bf Table}~\ref{tab1}.  No significant differences are found
with different evolutionary tracks.  The fit is designed to assign lower weight
to outlying points and to emphasize the bulk of stars near the blue border.
The distribution around the mean slope is well approximated by a Gaussian.
Standard deviations ($1\sigma$) of these distributions are listed in
parenthesis in Table~\ref{tab1}.

%**************************************
\begin{table}
\begin{center}
\caption{RR~Lyr fundamental blue edge slopes derived by different methods,
assuming a linear relation.}
\label{tab1}
\[
\begin{array}{ccc}
\hline
{\rm method} & {\rm slope} \ ($$1\sigma$$) & {\rm source}\\
\hline
\hline
{\bf empirical} & ${\bf -5.40} $ & {\rm Jurcsik} \ (\cite{jurcsik}) \\
\hline 
{\rm conv., lin.} & $-13.34 $ & {\rm KBF} \\
\hline
{\rm conv., \ nonlin., Dorman \ ev.} \ \ \  &  $-3.91 (0.51)$ \ \ & {\rm this \ paper} \ \\ 
\hline
{\rm conv., \ nonlin., Demarque \ ev.} \ \ \ &  $-3.96 (0.25)$ \ \ & {\rm this \ paper} \ \\ 
\hline
{\rm conv., \ nonlin., Padova \ ev.} \ \ \ &  $-4.25 (0.47)$ \ \ & {\rm this \ paper} \ \\ 
\hline
\end{array}
\]
\end{center}
\end{table}
%**************************************

%**************************************

\begin{figure*}
\includegraphics[width=18cm]{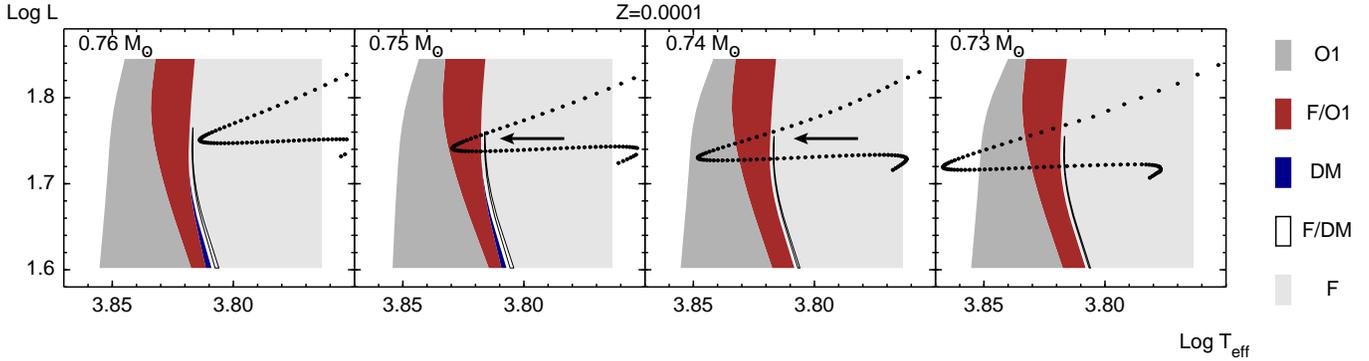}

\caption{Demarque (\cite{demarque}) evolutionary tracks superposed on modal
diagram for different masses for the case of low metallicity (Z=0.0001).  The
plotted points denoting the evolutionary tracks are uniformly distributed in
time. Color code is as in {\bf Fig.}~\ref{fig3}. F red edge is not shown.  {\bf Note
that for the sake of clarity the F/DM region has been plotted shifted by 100
${{\thinspace} {\rm K}}$ to the right in the Figures.}  Arrows indicate 
the occurrence of
double-mode pulsation.}

\label{fig5}
\end{figure*}
%**************************************

\subsection{The Results}

Our approach provides excellent agreement between empirical and theoretical F
blue edges in which previous simulations, whether radiative and convective,
completely failed, mainly because they did not take into account 
evolutionary effects. 
In fact the agreement in Fig.~\ref{fig4}. is almost perfect.
It is especially striking that the overall {\sl slope} of the blue edge of the
F mode perfectly matches its observed counterpart, and no significant
variations have been found for different evolutionary computations.  

We add a note of caution however about the agreement of the {\sl locations} of
the F blue edges. We recall that a tuning of the dimensionless
turbulent convection-related parameters results in small shifts.  Therefore the
almost perfect agreement for the location of the F blue edge may be a
coincidence, but we stress that {\sl the slope is invariant to such changes}.
  
Another achievement of these calculations is that the number density
distribution of stars along the blue edge bears a remarkable resemblance to
that of the empirical one ({\bf Fig.}~1. in KBF).  One notices a higher density at
higher $L$, and a sparser distribution to the lower luminosity regime.

It is important to call attention to the fact that {\it a priori\ } no distinct 
F blue edge is expected theoretically when stars with all kinds of metallicities 
and ages are included, and a well-mixed region is seen instead, as for Cepheids 
in the Magellanic Clouds (Udalski~{\it et~al.}  \cite{Udalski}).

From an observational point of view the F blue edge is equivalent to the
envelope of the RRab stars.  We emphasize that, because of the nature of the
evolutionary tracks, from the theoretical point of view the F blue edge
necessarily always means the blue edge of F/O1, and similarly, the O1 red edge
is defined by the red edge of the F/O1.  In {\bf Fig.}~\ref{fig4}. the F blue edge
provided by our method can also be defined as the blue envelope of the F
pulsators, formed through the joint effects of mode selection and evolution.

\subsection{Discussion}

In order to demonstrate the relevance of both mode selection and evolution in
blue edge modeling, we have carried out simple tests with
F/O1 edges of constant temperature, independent of ($L,M,Z$), as well as with
F/O1 edges with a linear dependence on L (and independent of $M,Z$). None
were successful in solving the F blue edge problem.
Therefore there was no choice but to carry out all the complicated,
time-consuming computations that we have just described.

Also, our earlier investigations (Szab\'o~{\it et~al.} \cite{rszabo1}) which
used model sequences of only one representative metallicity, did not alter the
slope significantly.  The motivation for this simple treatment was the huge
amount of computing capacity that was necessary to produce the diverse
metallicity-sequences. We note that earlier investigations (Szab\'o et
al. \cite{rszabo}) pointed out that the shape and the location of the DMR is
almost independent of $Z$, further complicating the question.  It is now clear,
that {\sl metallicity effects}  (and even the small temperature variation in IS
boundaries) also play role in the correct treatment of RR~Lyr ISs.

We have not considered the O1 blue edge here. Our method suggests a steeper
slope compared to the F blue edge, since no hysteresis mechanism acts here.
Considering O1 blue edges for fixed mass values, we can nicely reproduce 
other model computations (Bono~{\it et~al.} \cite{bono97}, KBF). 
Thus only special distributions along the evolutionary tracks (mass, metal content
and age distribution) can help to bend this slope in our framework, but we do not 
go into such speculations here. From an observational point of view there are too 
 few observed stars in this
region with reliable physical parameters.  This prevents any meaningful
comparison. Evidently future works should address this question, too.

%======================================================================

\section{Double-mode pulsation and evolution}
\label{sec6}

The characteristics of the DMR have been enumerated in the discussion of the
global mode selection map of Sec.~{\ref{sec4}}. Here we explore an additional 
aspect of DM pulsation, namely its connection with stellar evolution. We have mentioned
that the DMR is universally seen on the mode selection maps for all $Z$ and $L$
values.  On the other hand, it is well known that RRd stars are missing from
certain globular clusters (Clement~{\it et~al.} \cite{clement01}).  We propose
here that the mass distribution and stellar evolution may be the important
factors involved in the resolution of this puzzle.

The DMR for a sequence of mass with fixed metallicity ($Z=0.0001$) is presented in
the ${\rm Log}\thinspace {T_{\rm ef\!f} } - {\rm Log}\thinspace L$ diagram of
{\bf Fig.}~\ref{fig5}.
Note that again for better visibility the F/DM regions have been shifted 
100 ${{\thinspace} {\rm K}}$
to the right.  The bluest (interpolated) Demarque evolutionary tracks are also
displayed.  The arrows indicate where redward evolution cross the DMR, {{\it
i.e.}\ } where DM pulsation occurs. The Dorman
models are very close to the Demarque models, at slightly lower
temperatures, but the Padova tracks are much redder and do not touch our
computed DMRs.

%**************************************
\begin{figure*}
%\resizebox{\hsize}{!}{\includegraphics{pet2_col.ps}}
\centering
\includegraphics[width=9.2cm]{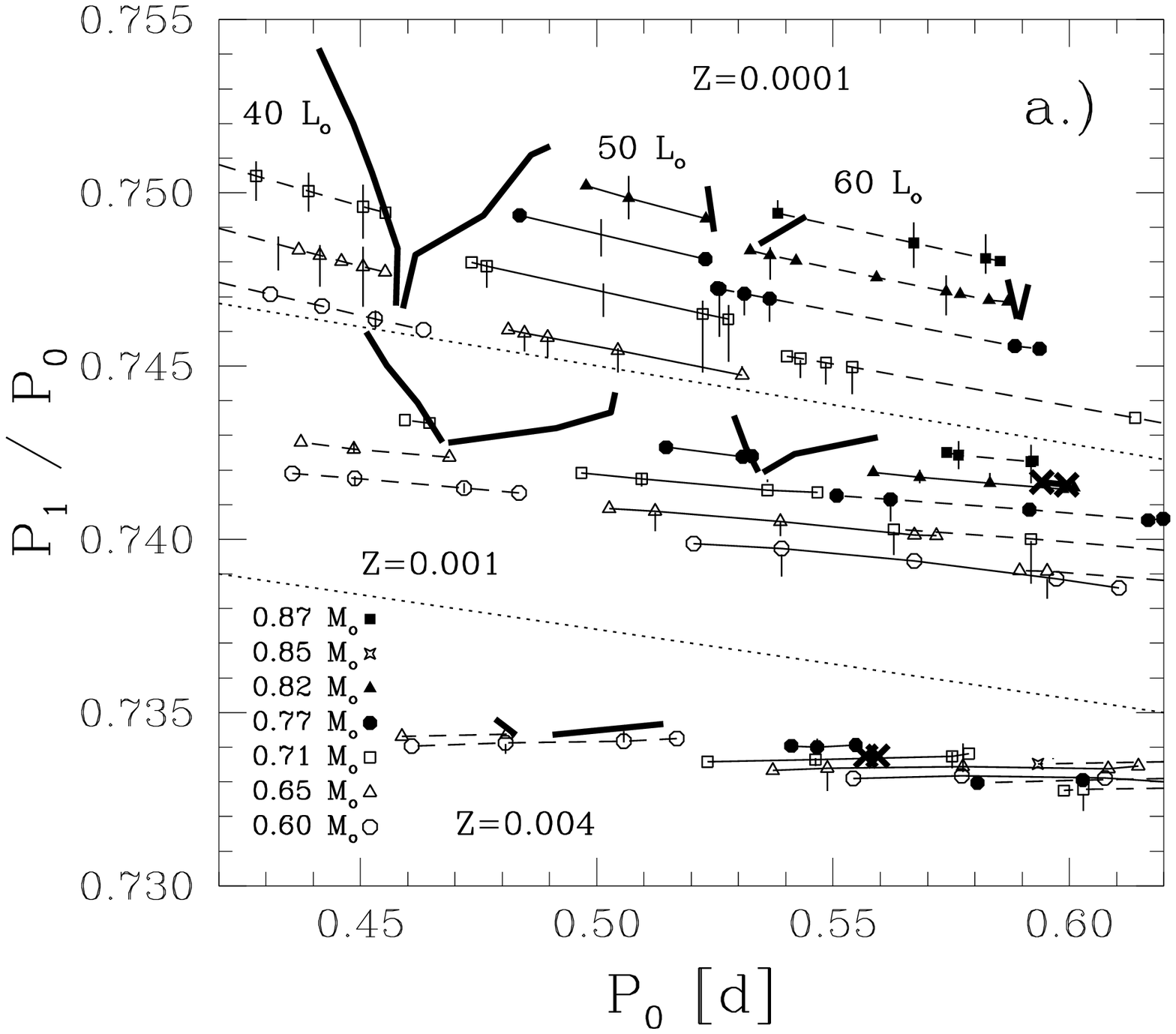}\includegraphics[width=9.2cm]{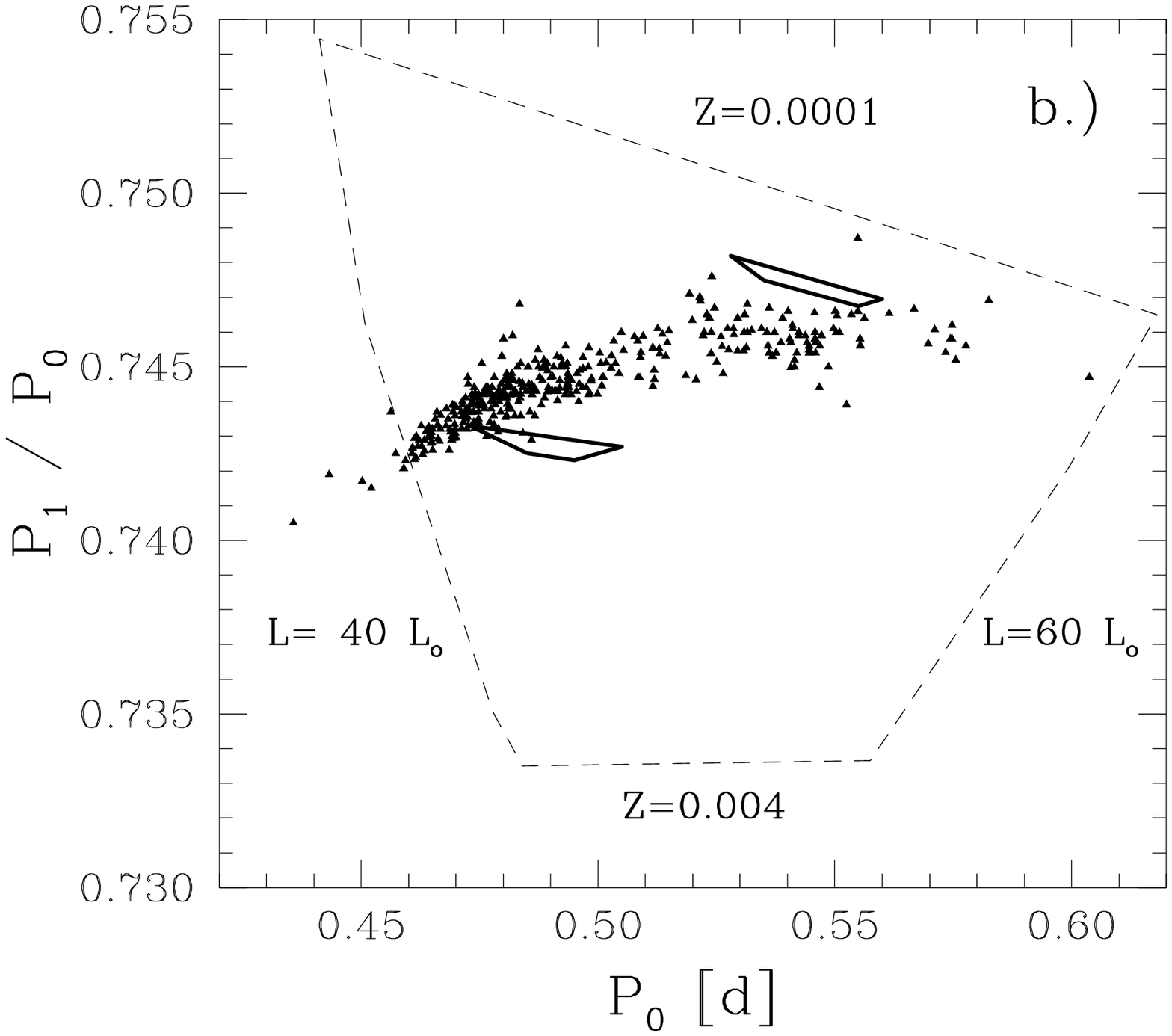}

\caption{{\bf a.)} F/O1 and DM period ratios for three luminosity and three
metallicity values derived form our models. Thin lines
are linear period ratios for F/O1 model sequences, ({{\it i.e.}} both F and O1 fixed
points are nonlinearly unstable). Nonlinear period shifts from F to O1 limit cycles 
(or vice versa) along the separatrix (see {\bf Fig}.~\ref{fig1}a) 
are displayed by short vertical bars. Double-mode
regions are represented by thick lines and big crosses. Models of different
metallicity are separated by dotted lines.
{\bf b.)} Petersen diagram of all known RRd stars denoted by small triangles.
The sources are the same as in Popielski~{\it et~al.}  (\cite{popielski}),
complemented by Sagittarius dwarf and Galactic Bulge RRd stars from
Cseresnjes (\cite{cseresnjes}) and Mizerski (\cite{mizerski}), as well as 
\object{LMC} and \object{SMC} stars
from Soszynski~{\it et~al.} (\cite{soszynski02}; \cite{soszynski03}) and
\object{Sculptor dSph} RRds from Kov\'acs (\cite{kovacs01}).  The extrema of ($Z,L$) that
delimit the possible DM pulsation region (taken from the left panel) are plotted
to guide the eyes (dashed lines). When evolution is considered, however, the 
DMRs are confined to the very small trapezoids: 
$Z=0.0001$ at upper right, $Z=0.001$ is lower left one.}

\label{fig6}
\end{figure*}
%**************************************

The most important feature of {\bf Fig.}~\ref{fig5}. is the narrow mass-range in which
DM pulsation is possible ($M=0.745 \pm 0.010\thinspace {\rm M_{\sun}}$). For 
$Z=0.001$ the mass range is similar, but it occurs for lower masses 
($M= 0.665 \pm 0.010 \thinspace {\rm M_{\sun}}$).  With $Z=0.004$ no tracks cross 
the DMR. This trend is in good agreement with the results of Popielski~{\it et~al.} 
(\cite{popielski}), even though they use different evolutionary and pulsational 
codes than we do.  Their method relies on investigating the evolutionary tracks and 
their mapping onto a Petersen diagram.  They obtain somewhat higher masses and a 
larger mass range ($0.79-0.855\thinspace {\rm M_{\sun}}$ with Z=0.0001, and 
$0.64-0.695\thinspace {\rm M_{\sun}}$ with $Z=0.001$, respectively).  But, we 
emphasize again that we are speaking about trends, and that altering the 
mode selection topology by fine-tuning the turbulent parameters can lead to a 
wider mass-range if the DMR goes to higher $L$ {{\it e.g.}}, or to an increase of 
the DM-mass by $0.01-0.02\thinspace {\rm M_{\sun}}$ if everything is
shifted to lower temperatures.

According to empirical evidence the metallicity of bulge RR~Lyrae ranges from 
$[Fe/H] = -1.5$ (0.0006) to $-0.5$ (0.006) (Walker \& Terndrup \cite{wt}). 
However the metallicity of double-mode RR~Lyrae does not necessarily follow this 
overall metallicity distribution: First,  bulge RRc stars do not show 
a high-metallicity tail (Walker \& Terndrup \cite{wt}), but a sharp cut around 
$[Fe/H] = -0.8$, and RRd stars presumably follow this behavior. Second, 
there is an immense contrast between the numbers of RRab/RRc and RRd stars: 
Cseresnjes (\cite{cseresnjes}) found 13 galactic RRd stars among 
$\sim 1850$ RR~Lyrae. Mizerski (\cite{mizerski}) has 
recently found only 3 RRd stars out of 2713 bulge RR~Lyrae based on OGLE-II 
observations. The small number of galactic bulge (and field) RRd stars inhibits 
the derivation of any direct connection between their metallicities and general 
RR~Lyrae metallicity distribution. Instead, this huge difference points to the natural 
selection effect we propose. Moreover, all the existing metallicity estimates or 
measurements of RRd stars tend to low values (Clementini~{\it et~al.} \cite{clementini}, 
Bragaglia~{\it et~al.} \cite{bragaglia}, hereafter BGC, Kov\'acs \cite{kovacs01}) and no 
bona fide high-metallicity RRd has been found to date, further supporting our findings.

We call attention to the fact that if the DMR consists of only a F/DM region
then solely redward evolution can produce DM pulsation.  Although we have
encountered this scenario we are not in the position to exclude the possibility
that blueward or both blue- and redward evolution produce DM stars.  Small
shifts in the locations of DM and F/DM regions or in the evolutionary tracks
may change the situation.  This clearly demonstrates the delicate interplay
between mode selection and evolutionary effects in determining the possible
parameter range of double-mode pulsation.

\subsection{Petersen diagram}
\label{petersen}

It is worth pointing out the fundamental difference between our theoretical Petersen 
diagram and previously published ones (Bono~{\it et~al.} \cite{bono96}, 
Bragaglia~{\it et~al.} \cite{bragaglia}, Popielski~{\it et~al.} 
\cite{popielski}). All the above works were based on F/O1 models in our
notation, whereas our Petersen diagram ({\bf Fig.}~\ref{fig6}b.) consists of only DM and 
F/DM models. To visualize the difference, we include {\bf Fig.}~\ref{fig6}a., where
both linear (and nonlinear) F/O1 and (nonlinear) DM period ratios are given without
evolutionary constraints.  Nonlinear period ratio shifts along the separatrix 
are also exhibited for selected models (see {\bf Fig}.~\ref{fig1}a). From F to O1 
$P_{1} / P_{0}$  increases, from O1 to F decreases compared to its linear
counterpart. These values can be important for mode switching stars at both 
ends of the F/O1 region. Note that DMR always follows F/O1
region at the longer period (lower temperature) side, and at 
higher masses (cf. {\bf Fig}.~\ref{fig3}). 

The Petersen diagram supports our combined DM-pulsational and evolutionary 
considerations. We have plotted all the published period 
ratios of RRd stars in {\bf Fig.}~\ref{fig6}b.  Mode selection in itself does
not constrain the possible DM region as seen on panel a.) and also from the position of 
the dashed lines on panel b.). These lines are plotted solely for orientation, as extending the 
luminosity and metallicity range would further expand the permitted region. However, if
evolution is taken into account, {{\it i.e.}} we only consider the tracks
that cross the DMR, then the permitted region is approximately confined to the
observed distribution on the Petersen diagram. As we have merely two
metallicity values ($Z=0.0001, \ 0.001$) that are approximated by the two lozenges, we
cannot reproduce the bend of the RRd star distribution, but our results are close
to the observed features, though not perfect yet. We emphasize that this is the first 
attempt at reproducing the RR~Lyr Petersen diagram on the basis of nonlinear DM models.

Another difference lies in the use of DM models only for RRd stars instead of
non-DM models. Namely, as these models 
can be found in a much narrower region than F/O1 counterparts ({\bf Fig.}~\ref{fig3}), 
therefore their transformation to the Petersen diagram is also much more localized. This 
effect is intensified by evolutionary selection effects that lead to a shrinking of 
the possible DM areas, as seen in {\bf Fig.}~\ref{fig6}b. Thanks to the application of
DM models and the inclusion of the evolutionary considerations, the net effects 
are: a.) the lifting of the persistent, disturbing mass-metallicity degeneracy and
b.) the emergence of a more consistent picture, where evolutionary constraints force
period ratios of the nonlinear DM-models to stretch along the observed distribution of
RRd stars. 

A comparison can be made with BGC published $Z=0.0001$ models (their Table~6.), which 
is an extension of the work described in Bono~{\it et al.} (\cite{bono96}). 
Our F/O1 periods are quite similar (and almost parallel) to the BGC 
sequences. The location of our nonlinear DM (both DM and F/DM) 
models are somewhat different from that of the BGC models of the same $(L,M,Z)$ parameters: 
the $Z=0.0001$ trapezoid is at higher period ratio. The discrepancy may partly be 
ascribed to the fact that F/O1 regions are replaced by DMR at high masses, so the 
overall picture is different. A slightly smaller period ratio (by $0.001-0.002$) at this
metallicity (similar that of BGC) would mean a better agreement with the observed
distribution. The discussed possible shifts of the modal selection structures to 
higher $L$ and/or alteration of evolutionary tracks in the same sense would shift 
our region to the lower right, again resulting in an almost perfect explanation of
the high-metallicity distribution of the observed RRd sample. In the $Z=0.001$ case 
either our RRd masses are too low, or a DM region being at higher temperature 
instead of F/DM would yield closer agreement between the observations and our predictions. 
Currently none of these explanations are supported by our models.

Feuchtinger (\cite{feuchtinger1}) published the first nonlinear DM model ($M=0.65
M_{\sun}$, ${\rm Log}\thinspace L = 1.72$, $Z=0.001$, $T_{\rm ef\!f} = 6820 K$, 
$P_{0} = 0.5279$, $P_{1} / P_{0}= 0.7468$). His model is at higher temperature, but the 
mass value perfectly agrees with ours, though no evolutionary effects were considered. 
Its position on the Petersen diagram disagrees with the linear and nonlinear single mode 
calculations and with the metallicities of the observed RRds in this region. It lies between  
our  $Z=0.001$ and $Z=0.0001$ areas. Thus our results 
are closer to earlier works based on F/O1 models, such as Bono~{\it et~al.} 
(\cite{bono96}), Bragaglia~{\it et~al.} (\cite{bragaglia}), Popielski~{\it et~al.} 
(\cite{popielski}). The only possible difference between our and Feuchtinger's code 
producing nonlinear DM models is a different turbulent convection parameter set. 
The astrophysical calibration of these parameters is still not a settled 
question as discussed several times throughout this paper. Interestingly, this 
discrepancy suggests the use of the Petersen diagram itself for the
purpose of calibrating the  convective parameters. 
An additional constraint for the models can be to ensure a good agreement with the 
the observed distribution on the $P_{0} - P_{1} / P_{0}$ plane, while simultaneously 
respecting other constraints, too (amplitudes, IS width etc.). Because 
of the nontrivial nature of this task and the huge amount of necessary computational 
time, this question is not addressed in the present work. A future work will be 
initiated to clear this situation.

\vskip15pt

To sum up this section, the following factors should be kept in mind when
explaining the wealth of RRd stars in certain globular clusters (\object{IC4499},
\object{M68},
etc.), the low number of RRd stars in the Galactic field and other globular
clusters ({{\it e.g.}\ } \object{M3}) or even their complete absence despite
the considerable abundance of RR~Lyr stars:

\noindent -- a very narrow mass range at a given metal content,

\noindent -- a small temperature range, consequently a limited time-span for a given star,

\noindent -- no high $Z$ evolutionary tracks crossing DMR.

We think that while our work is an important step in the right direction,
much work remains to be done before a complete self-consistent
scenario can emerge.

%======================================================================

\section{Conclusions}
\label{sec7}

We have developed a very efficient methodology for studying the evolution of the
pulsations of a given stellar model along its evolutionary track through the
IS.  It involves a judicious mixture of numerical hydrodynamics,
analytical signal time-series analysis, and amplitude equations.

With the help of nonlinear, turbulent convective model computations we have
analyzed in detail the modal topology of the RR~Lyr IS.  We have
provided data on the F and O1 pulsation regimes as well as on DM regions as a
function of metallicity and stellar mass.  Details of the important problem of
mode selection have been presented.  We note that our methodology allows us to
find and delineate the very narrow DM regimes very effectively, while, in
contrast, with numerical hydrodynamic studies this would be akin to looking for
a needle in a haystack.

Synthetic ${T_{\rm ef\!f} }-L$ diagrams of F and O1 RR~Lyr stars have been
generated.  Toward this end we have carried out Monte-Carlo simulation for
three different sets of evolutionary tracks with the purpose of deriving the
RR~Lyr F blue edge.  The combination of numerical modeling of stellar pulsation
including turbulent convection and evolutionary calculations reproduces the
linear shape of the F blue edge despite the complex topology of the mode
selection.  We obtain essentially agreement between the empirical and the
theoretical slopes, as well as for the density of stars in the IS, thus
resolving an extant puzzle (KBF).  It is found that observed structure of
the IS can only be accounted for with a correct treatment of
metallicity, mass, evolutionary effects, and nonlinear mode selection modeling.
Our results provide a new, independent constraint on horizontal-branch
evolution, and confirm the results of theoretical evolutionary calculations.

Based on our calculations with a large number of stars of different
metallicities and masses,  we find that the RRab/RRc populations are not well
separated, and that there should be an area in the 
${\rm Log}\thinspace {T_{\rm ef\!f} } - {\rm Log}\thinspace L$ diagram that contains 
both RRab and RRc stars (sometimes referred to as an either-or region).

A strong (anti-)\thinspace correlation is demonstrated between the F/O1 and DM 
regimes. A DM region occurs at high mass, while at low mass a F/O1 region is 
invariably found.  The transition mass between the two features depends on 
luminosity. Hence the simple statements that (1) RRd stars lie between the 
RRab and RRc pulsators, and that (2) F/O1 is between F and O1 region, can be 
quite misleading.

RRd stars are observed in a very narrow range of parameters (color index,
temperature). On the basis of our survey we find that the possible region of
DM behavior is indeed very narrow.  Furthermore we show that evolution selects an even
smaller region, {\it viz.\ } a mass-range of extremely small width 
($0.02\thinspace {\rm M_{\sun}}$) at $Z=0.0001 - 0.001$. The evolutionary tracks do not 
cross the DMR regime for high $Z$. The computed period ratios are in reasonable agreement 
with the observational Petersen diagram, provided that stellar evolution is taken into account. 
The small existing discrepancy may be ascribed to uncalibrated factors in the
description of the turbulent convection, but in turn may be used to determine them. 

As an extension of this work, a
detailed analysis with the goal of deriving empirical relations between
light-curve and physical properties of RRc stars would be of great value.
Reliable light-curves (and also color-indices) of RR~Lyr populations of
LMC/SMC or nearby dwarf galaxies that have emerged or will
emerge from large-scale surveys are bound to extend the validity of and improve
the agreement between empirical and theoretical results of IS
investigations.

This survey did not specifically search for second overtone pulsations, but we
note that we have not encountered any RRe models.  But to settle this issue, a
special survey will need to be made.

%**************************************
\begin{acknowledgements}

This work has been supported by NSF (grant AST03-07281) and the Hungarian OTKA
(T-038440 and T-038437). Fruitful discussions with J. Jurcsik and G. Kov\'acs 
are appreciated. We thank the anonymous referee for valuable comments and suggestions 
which helped to improve the paper. 

\end{acknowledgements}
%**************************************

%************************************** 

\end{document}